\documentclass[letterpaper]{article} % DO NOT CHANGE THIS
\usepackage[]{aaai2026}  % DO NOT CHANGE THIS
\usepackage{times}  % DO NOT CHANGE THIS
\usepackage{helvet}  % DO NOT CHANGE THIS
\usepackage{courier}  % DO NOT CHANGE THIS
\usepackage[hyphens]{url}  % DO NOT CHANGE THIS
\usepackage{graphicx} % DO NOT CHANGE THIS

\usepackage[table]{xcolor} % Added for \colorbox
\usepackage{booktabs}
\usepackage{multirow}

\usepackage{natbib}  % DO NOT CHANGE THIS AND DO NOT ADD ANY OPTIONS TO IT
\usepackage{caption} % DO NOT CHANGE THIS AND DO NOT ADD ANY OPTIONS TO IT
\usepackage{algorithm}
\usepackage{algorithmic}
\usepackage{amsmath}
\usepackage{amsfonts}
\usepackage{newfloat}
\usepackage{listings}
\DeclareCaptionStyle{ruled}{labelfont=normalfont,labelsep=colon,strut=off} % DO NOT CHANGE THIS
\floatstyle{ruled}
\newfloat{listing}{tb}{lst}{}
\floatname{listing}{Listing}
\title{Laplacian Analysis Meets Dynamics Modelling: Gaussian Splatting \\ for 4D Reconstruction}
\author {
    Yifan Zhou \thanks{These authors contributed equally to this work.}  \textsuperscript{\rm 1},
    Beizhen Zhao \footnotemark[1]  \textsuperscript{\rm 1},
    Pengcheng Wu \textsuperscript{\rm 2},
    Hao Wang  \thanks{Corresponding author.}  \textsuperscript{\rm 1}
}
\affiliations {
    \textsuperscript{\rm 1} ‌The Hong Kong University of Science and Technology (Guangzhou) \\
    \textsuperscript{\rm 2} Nanyang Technological University\\
    yifanzhou@hkust-gz.edu.cn, bzhao610@connect.hkust-gz.edu.cn, pengchengwu@ntu.edu.sg, haowang@hkust-gz.edu.cn
}

\usepackage{bibentry}
\begin{document}

\maketitle

\begin{abstract}
While 3D Gaussian Splatting (3DGS) excels in static scene modeling, its extension to dynamic scenes introduces significant challenges.
Existing dynamic 3DGS methods suffer from either over-smoothing due to low-rank decomposition or feature collision from high-dimensional grid sampling. 
This is because of the inherent spectral conflicts between preserving motion details and maintaining deformation consistency at different frequency. 
To address these challenges, we propose a novel dynamic 3DGS framework with hybrid explicit-implicit functions. 
Our approach contains three key innovations: 
a spectral-aware Laplacian encoding architecture which merges Hash encoding and Laplacian-based module for flexible frequency motion control, 
an enhanced Gaussian dynamics attribute that compensates for photometric distortions caused by geometric deformation,
and an adaptive Gaussian split strategy guided by KDTree-based primitive control to efficiently query and optimize dynamic areas.
Through extensive experiments, our method demonstrates state-of-the-art performance in reconstructing complex dynamic scenes, achieving better reconstruction fidelity.

\end{abstract}

\section{Introduction}

Dynamic scene reconstruction from monocular videos presents a critical challenge in computer vision, demanding precise modeling of both persistent geometric structures and transient deformations~\cite{cai2022neural,du2021neural,fang2022fast,kratimenos2024dynmf,li2022neural}. 
Unlike static environments, dynamic scenes exhibit heterogeneous motion patterns - rigid components maintain temporal consistency while deformable regions require high-frequency trajectory modeling~\cite{xu2024grid4d,duan20244d,bae2024per}. 
This inherent complexity creates dual challenges: preserving spatial coherence across time-varying geometries and capturing transient deformation details without over-smoothing artifacts.

While Neural Radiance Fields (NeRF)~\cite{mildenhall2021nerf, chan2022efficient, yang2022banmo,park2021hypernerf, martin2021nerf} revolutionized static scene modeling through continuous volumetric integration, their dynamic extensions~\cite{choe2023spacetime, gao2021dynamic, liang2023semantic, liu2023robust, wang2023flow, barron2021mip, fridovich2023k} reveal critical limitations in handling temporal discontinuities, particularly the conflicting requirements for spatial fidelity versus temporal coherence arising from uniform spectrum allocation. 
Although explicit representations~\cite{barron2023zip, cao2023hexplane, tancik2022block} improve efficiency through 4D spacetime factorization, their low-rank decomposition induces feature collision in overlapping regions. 
Recent 3D Gaussian Splatting (3DGS)~\cite{kerbl20233d, duisterhof2023md, yang2023real, liang2023gaufre,lin2024gaussian} has achieved impressive effects for static environments, where discrete volumetric primitives enable both photorealistic rendering and computationally efficient optimization through differentiable rasterization~\cite{shao2023tensor4d, wu20244d,lu20243d,luiten2024dynamic,kratimenos2024dynmf}. 

However, their direct extension to dynamic scenarios faces three fundamental limitations: 
1) existing deformable methods suffer from either over-smoothing due to low-rank decomposition or feature collision from high-dimensional grid sampling, 
2) previous Gaussian-based methods use a fixed threshold during Gaussian split stage which ignore adaptive split adjustment, and 
3) persistent appearance changes caused by dynamic deformation are often neglected in current pipelines.

To address the challenges above, our key insight lies in addressing the anisotropic spatio-temporal sampling nature of dynamic scenes through hybrid explicit-implicit encoding.
First, we develop a hybrid spectral-aware encoder combining Hash grids with Laplacian-based module that decouples spatial and temporal features into different frequency motion components, overcoming the feature collision of low-rank assumption while enabling adaptive frequency motion control. 
Then, we design an enhanced Gaussian dynamics attribute to perform individual Gaussian personalized dynamic optimization and design an adaptive regularization for identifying highly dynamic areas. 
Besides, we propose an adaptive Gaussian split strategy, which focuses on the optimization trade-off between Gaussian shape and anisotropy in dynamic scenes and an improved KDTree-based clustering algorithm was proposed to efficiently query and optimize dynamic Gaussians.

Our solution rethinks dynamic 3DGS through Laplacian spectral analysis, which provides a hybrid framework for localized frequency analysis. 
Meanwhile, we focus on the dynamics attribute of each Gaussian and the optimization problem in the derivation process, and propose a novel hybrid explicit-implicit algorithm model.
We propose three key innovations.
In summary, our contributions are as follows:
\begin{itemize}
    \item We propose a spectral-aware Laplacian encoding module combining multi-scale Hash encoding with Laplacian-based dynamic module that decouples different frequency motion trajectories from complex deformation.
    \item We design an enhanced Gaussian dynamics attribute that identify highly dynamic areas for adaptive split and regularization.
    \item We design an adaptive Gaussian split strategy that automatically adjusts the primitive density and anisotropy using KDTree-guided spectral analysis.
\end{itemize}

\section{Related Work}

\subsection{NeRF-based Dynamic Modeling}

The advent of Neural Radiance Fields (NeRF) has significantly transformed the landscape of 3D scene reconstruction, particularly for static environments. 
However, extending NeRF to effectively model dynamic scenes remains a formidable challenge. 
Initial efforts, such as D-NeRF~\cite{pumarola2021d} and Nerfies~\cite{park2021nerfies}, have employed canonical space warping and temporal latent codes to capture motion. 
Despite their innovative approaches, these methods often exhibit limitations when dealing with rapid or abrupt movements.
Moreover, explicit spacetime factorization techniques, such as HexPlane~\cite{cao2023hexplane}, have been proposed to enhance computational efficiency. 
However, these methods impose restrictive low-rank assumptions that may oversimplify the intricate dynamics present in real-world scenes, particularly in environments characterized by rapid changes.
Furthermore, while segmenting scenes into components with distinct attributes has been explored to enhance modeling accuracy~\cite{gao2021dynamic, tretschk2021non}, the implicit representations based on fully connected MLPs often suffer from over-smoothing and lengthy training processes. 

\subsection{3DGS-based Dynamic Modeling}

Recent advances in 3D Gaussian Splatting (3DGS)~\cite{kerbl20233d,yu2024cogs,huang2024sc, li2024spacetime} have demonstrated remarkable success in static scene reconstruction, prompting extensions to dynamic scenarios. 
While 4D-GS~\cite{wu20244d} employs multi-resolution HexPlanes with MLPs for deformation modeling, it inherits the fundamental limitation of plane-based methods: the low-rank assumption leads to feature collisions and rendering artifacts in complex motions. 
Neural deformation fields~\cite{yang2024deformable,huang2024sc} address this through MLPs, but often produce over-smoothed results and struggle with high-frequency details due to insufficient inductive biases. 
Direct optimization of 4D Gaussians~\cite{yang2023real,duan20244d} offers greater flexibility but introduces optimization challenges including floating artifacts and requires extensive training with additional regularizers. 
Grid4D~\cite{xu2024grid4d} has achieved impressive performance through combining triplane and Hash-coding while it often lacks smoothness and works without an explicit method for modeling dynamic processes.  
While SplineGS~\cite{park2024splinegs} proposes a pipeline that combine 3DGS and spline functions, however, it requires massive priors such as 2D trajactory and depth estimation to maintain performance.
These limitations collectively highlight the need for a representation that balances expressiveness with efficient optimization for dynamic 3DGS, particularly in handling complex motions while preserving fine details.
Our work addresses these limitations through a novel Laplacian motion representation that jointly optimizes for physical plausibility, computational efficiency, and multi-scale temporal fidelity.

\begin{figure*}[t]
\centerline{\includegraphics[width=1.0\textwidth]{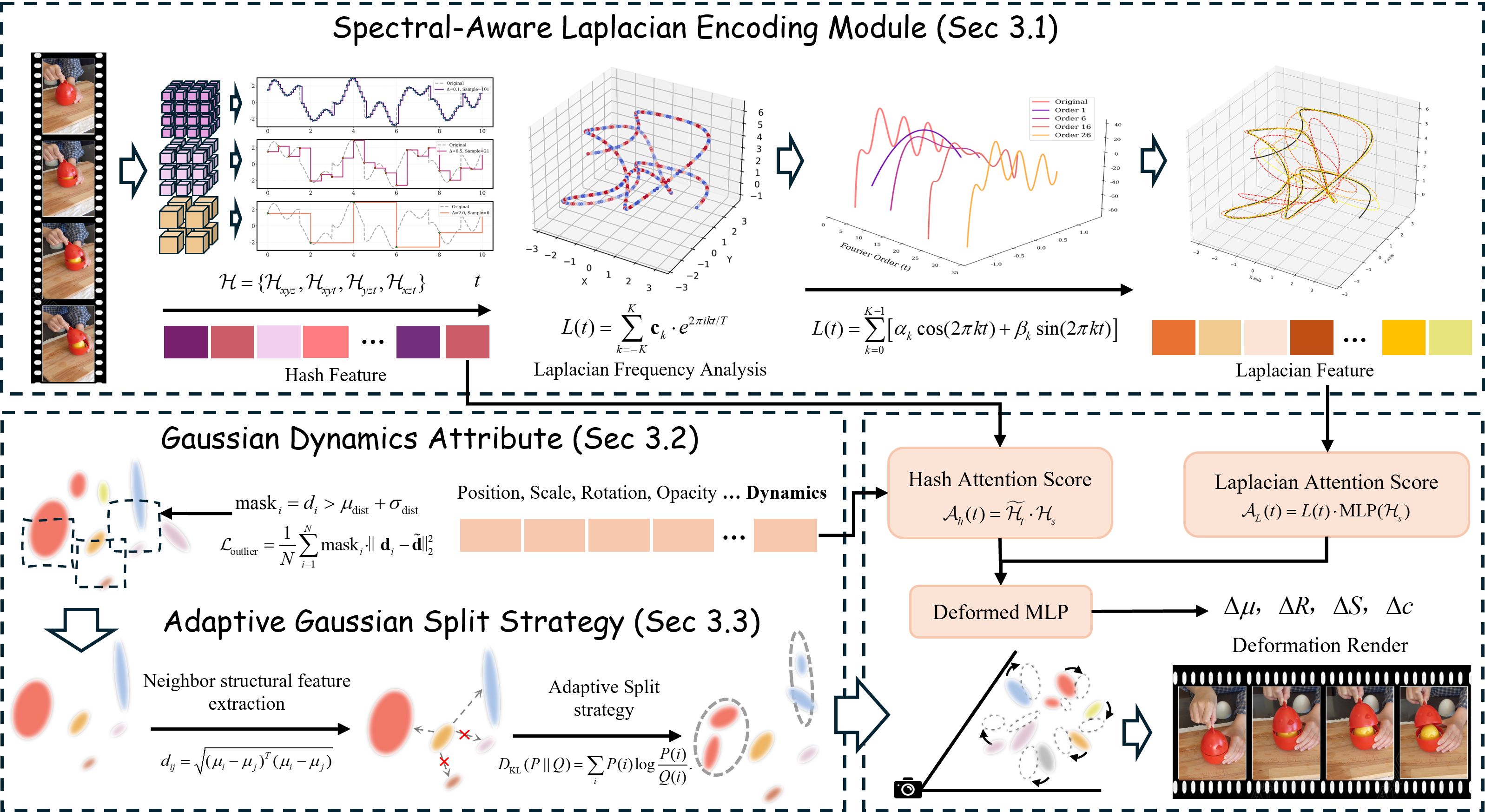}}
\caption{\textbf{Framework of our method.} We begin with a multi-scale Hash encoder to extract spatio-temporal features of Gaussians. Then a hybrid laplacian-based dynamic module is designed for adaptive frequency analysis. We design an appearance adjustment module which combines the Gaussian dynamic attribute with the spatio-temporal feature by attention mechanism. The whole pipeline benefits from the adaptive dynamic Gaussian split strategy for better performance and efficiency. }
\label{fig1}
\end{figure*}

\section{Methodology}

In this section, we present our methodology aimed at addressing the challenges of modeling 4D dynamic scenes with high-fidelity spatial details and complex temporal variations.
The key innovation lies in a hybrid explicit-implicit representation that combines multi-scale Hash encoding with spectral decomposition to capture spatial features and adaptive temporal dynamics.
This framework is structured into three main components: spectral-aware Laplacian encoding module, enhanced Gaussian dynamic attribute and adaptive Gaussian split strategy. 
This hybrid approach is designed to enhance the representation of motion dynamics while maintaining physical consistency across spatial and temporal domains, significantly outperforming existing methods in handling complex motion patterns.
The overall pipeline is shown in Fig. \ref{fig1}.

\subsection{Spectral-Aware Laplacian Encoding Module}

This section focuses on the challenge of effectively encoding spatial and temporal information to capture the dynamics of motion. 
We employ a spectral-aware Laplacian encoding module that decomposes the frequency motion trajectories to accommodate the complexities of 4D spacetime.

\subsubsection{Multi-Scale Hash Encoding}
To efficiently encode 4D spacetime information while preserving both spatial and temporal details, we employ a multi-scale hash encoding strategy that extends traditional methods.
Inspired by Grid4D~\cite{xu2024grid4d}, we extend InstantNGP's Hash encoding~\cite{muller2022instant} to 4D spacetime $(x,y,z,t)$ through anisotropic multi-resolution decomposition.

\begin{equation}
    \mathcal{H}^l = \{\mathcal{H}_{xyz}^l, \mathcal{H}_{xyt}^l, \mathcal{H}_{yzt}^l, \mathcal{H}_{xzt}^l\}, \quad l \in \{1,...,L\}
\end{equation}

Each level $l$ maintains dimension-specific resolutions computed via geometric progression.

\subsubsection{Laplacian-Based Motion Prediction}

In the realm of dynamic scene reconstruction, accurately predicting motion dynamics is paramount. 
Traditional methods often rely on MLP or linear interpolation, which fails to capture the complex periodic and aperiodic motions present in real-world dynamic scenes. 
To overcome these limitations, we propose a novel hybrid explicit-implicit Laplacian-based motion representation that combines the expressiveness of spectral analysis learnable neural components, allowing for the effective capture of both low and high-frequency motion dynamics.
The foundation of our approach lies in the Laplacian series decomposition of time series:

\begin{equation}
    L(t) = \sum_{k=-K}^K \mathbf{c}_k \cdot e^{2\pi ikt/T},
\end{equation}
where $L(t)$ represents the Laplacian motion field at time $t$, with $\mathbf{c}_k$ denoting coefficients. 
Through Euler's formula, we substitute this representation into our motion prediction:

\begin{equation}
    L(t) = \sum_{k=-K}^K \mathbf{c}_k \cdot \cos\left(\frac{2\pi kt}{T}\right) + i \sum_{k=-K}^K \mathbf{c}_k \cdot \sin\left(\frac{2\pi kt}{T}\right).
\end{equation}

This transformation naturally handles periodic motions common in dynamic scenes and the frequency components provide interpretable control over motion characteristics.
To further enhance our motion representation, we extend the equation to a simple formulation:

\begin{equation}
    L(t) = \sum_{k=0}^{K-1} \left[ \alpha_k \cos(2\pi k t) + \beta_k \sin(2\pi k t) \right].
\end{equation}

Here, coefficients $(\alpha_k, \beta_k)$ are learnable parameters. 
This design aims to automatically adapt to scene-specific motion frequencies while maintaining end-to-end differentiability. 
The Laplacian-based decoding allows our model to learn appropriate frequency compositions directly from data, eliminating the need for manual frequency band selection.

The orthogonal basis properties enable stable gradient computation during optimization.
The incorporation of learnable frequencies $f_k$ through gradient enhances the ability to capture different frequency motion components: 

\begin{equation}
\frac{\partial \mathcal{L}}{\partial f_k} = \frac{1}{\sigma_k^2} \sum_{t} \left( \frac{\partial \mathcal{L}}{\partial L(t)} \cdot t \cdot [-\alpha_k \sin(2\pi f_k t) + \beta_k \cos(2\pi f_k t)] \right),
\end{equation}
where $\sigma_k$ denotes temporal variance. 
This mechanism automatically balances frequency preservation with motion stability. 
Then we introduce an attention mechanism which combines Laplacian features with Hash spatial features $\mathcal{H}_{s}$:

\begin{equation}
    \mathcal{A}_L(t) = L(t) \cdot \text{MLP}(\mathcal{H}_{s}).
\end{equation}

\subsubsection{Multi-Scale Laplacian Pyramid Supervision}
To enforce consistency across different frequency bands and spatial scales, we introduce a multi-scale supervision strategy that enforces consistency across frequency bands, enhancing the model's ability to better detail preservation.
We supervise reconstruction using Laplacian pyramid decomposition:

\begin{equation}
    \mathcal{L}_{\text{lap}} = \sum_{l=1}^L \lambda_l \|\mathcal{L}_l(I_{\text{render}}) - \mathcal{L}_l(I_{\text{gt}})\|_1,
\end{equation}
where $\lambda_l$ decreasing exponentially to emphasize finer details. 
This loss function encourages the model to focus on both coarse and fine features, ensuring a comprehensive understanding of motion dynamics.

\subsection{Enhanced Gaussian Dynamics Attribute with Adaptive Regularization}
\label{sec:dynamic_attributes}

To effectively model the dynamic variations inherent in complex scenes, we augment the standard 3D Gaussian Splatting representation. 
Specifically, we associate each 3D Gaussian $G_i$ with a learnable dynamics attribute, denoted as $\mathbf{d}_i \in \mathbb{R}^{D_d}$, where $D_d$ represents the dimensionality of this attribute space.
The dynamics attribute $\mathbf{d}_i$ is introduced to explicitly encapsulate these latent per-Gaussian temporal or conditional variations, providing a dedicated representation for dynamic properties.

To further improve the modeling of dynamic scene changes, we introduce a fusion mechanism that concatenates the original dynamic attribute vector $\mathbf{d}_i$ with the Hash temporal feature $\mathcal{H}_{t}$, forming an augmented feature representation:

\begin{equation}
    \tilde{\mathcal{H}_{t}} = \text{Concatenate}(\mathbf{d}_i, \mathcal{H}_{t}).
    \label{eq:fusion}
\end{equation}

This concatenation provides a straightforward yet effective means of integrating scene-specific temporal information with the Gaussian's intrinsic dynamic attributes, enabling the model to leverage both sources for more accurate deformation prediction.
To effectively combine spatial and temporal information, we introduce an attention mechanism to aggregate spatio-temporal features through:

\begin{equation}
    \mathcal{A}_h(t) = \tilde{\mathcal{H}_{t}} \cdot \mathcal{H}_{s}.
\end{equation}

\subsubsection{Adaptive Dynamic Regularization}
To ensure that our method better model dynamic changes, we implement a selective regularization mechanism that targets only those Gaussians exhibiting "abnormally large" or "highly dynamic" changes. 
These gaussians are referred to as "outliers", which need to be increase their gradients and thus promote their deformation or dynamic transformations. 

Instead of using fixed thresholds or applying a regularization on all gaussians, our method employs a data-driven, adaptive dynamic selection scheme. Specifically, for each Gaussian, we compute the Euclidean distance between its dynamic attribute $\mathbf{d}_i$ and a reference mean dynamic attribute $\mathbf{\bar{d}}$, as well as the associated standard deviation:

\begin{equation}
    d_i = \|\mathbf{d}_i - \mathbf{\bar{d}}\|_2.
\end{equation}

Let $\mu_{\text{dist}}$ and $\sigma_{\text{dist}}$ denote the mean and standard deviation of all $\text{d\_dist}_i$ across the Gaussian set. We then generate a mask to identify those points that are significantly deviating from the typical embedding variation:

\begin{equation}
    \text{mask}_i = d_i > \mu_{\text{dist}} + \sigma_{\text{dist}}.
\end{equation}

Only the Gaussians satisfying this outlier criterion—i.e., those with $\text{mask}_i = 1$—are subjected to the additional regularization loss:

\begin{equation}
    \mathcal{L}_{\text{dy}} = \frac{1}{N} \sum_{i=1}^{N} \text{mask}_i \cdot \|\mathbf{d}_i - \mathbf{\bar{d}}\|_2^2.
    \label{eq:outlier_loss}
\end{equation}

The purpose of this selective regularization is to intensify the gradients for Gaussians exhibiting large changes, thereby explicitly promoting their deformation and densification. 
By employing this dynamic regularization mechanism, the model adaptively concentrates regularization efforts on the most informative and dynamically relevant regions, effectively enhancing the capacity to model complex scene dynamics without imposing uniform constraints across all Gaussians.

In addition, we use Normalized cross-correlation (NCC)~\cite{yoo2009fast} based loss function $\mathcal{L}_{\text{NCC}}$ to evaluate the similarity between two images within local regions while maintaining invariance to brightness variations to enhance the alignment accuracy.
The total loss $\mathcal{L}$ used for training is composed of four distinct loss terms, each weighted by a corresponding hyperparameter $\lambda$ to control its relative contribution. 
The total loss is formulated as:

\begin{equation}
    \mathcal{L} = \mathcal{L}_{\text{orig}} + \lambda_{\text{NCC}} \mathcal{L}_{\text{NCC}} + \lambda_{\text{lap}} \mathcal{L}_{\text{lap}} + \lambda_{\text{dy}} \mathcal{L}_{\text{dy}},
    \label{eq:total_loss}
\end{equation}

where $\mathcal{L}_{\text{orig}}$ denotes original loss function of 3DGS and consists of $\mathcal{L}_\mathrm{1}$ and Structural Similarity Index Measure (SSIM) loss functions~\cite{wang2004image} $\mathcal{L}_\mathrm{SSIM}$.

\subsection{Adaptive Gaussian Split Strategy}

In this section, we address the challenge of optimizing Gaussian representations of motion dynamics. 
Our approach utilizes the analysis about the structure of each Gaussian to adaptively refine Gaussian parameters based on local neighborhood information, enhancing the model's ability to capture complex motion patterns.

\subsubsection{KDTree-Based Primitive Analysis} 
To maintain spatial coherence and prevent overfitting, we analyze Gaussian primitives through their neighborhood relationships.
For each Gaussian $\mathcal{G}_i$, we find $K$ nearest neighbors $\{\mathcal{G}_j\}_{j=1}^K$ based on Euclidean distance:

\begin{equation}
    ed_{ij} = \sqrt{(\mu_i - \mu_j)^T(\mu_i - \mu_j)}.
\end{equation}

By examining the size and anisotropy of each Gaussian ~\cite{xie2024physgaussian}, we can determine which Gaussians exhibit significant differences in their motion characteristics. 
Covariance differences are computed through L2 norm:

\begin{equation}
    \Delta\Sigma_{ij} = \|\Sigma_i - \Sigma_j\|_2.
\end{equation}

This adaptive approach ensures that our models remain responsive to local variations in motion dynamics.
By focusing on Gaussians with notable differences in size and anisotropy, we can selectively choose which Gaussian to split, thereby enhancing the model's ability to represent complex motion patterns without introducing unnecessary complexity.

\subsubsection{KL-Divergence Guided Adaptation}
In some cases, we observed that the KDTree-based partitioning method can not accurately identify the dynamic Gaussian, which leads to the instability to capture dynamic motions.
This is because the strategy of splitting Gaussian based on hard threshold will stop deriving when the shape and size of Gaussian in the neighborhood are similar. 
However, it is possible that some Gaussian in the neighborhood still contribute to the dynamic modeling.
To further refine the Gaussian split process, we compute the KL-divergence between the neighbor Gaussian distribution $P$ and a uniform distribution $Q$:

\begin{equation}
    D_{\text{KL}}(P \parallel Q) = \sum_{k=1}^K P(k)\log\frac{P(k)}{Q(k)}
\end{equation}

The adaptive splitting threshold $\tau$ becomes:

\begin{equation}
    \tau = \Delta\Sigma + D_{\text{KL}} \cdot \tau_{\text{base}}
\end{equation}

where $\tau_{\text{base}}$ is a hyperparameter. 
This mechanism allows for dynamic adjustments to the model's complexity based on the observed motion patterns.
Through adaptive dynamic Gaussian optimization strategy, we can determine when a Gaussian should be split more effectively, ensuring the model captures the nuances of motion dynamics while maintaining computational efficiency. 
This strategy enhances the model's robustness against overfitting by focusing on the most relevant Gaussian structures.

\begin{table*}[t]
\caption{\textbf{Quantitative comparison to previous methods on HyperNeRF ~\cite{park2021hypernerf} dataset.} The higher PSNR(↑) and higher SSIM(↑) denote better rendering quality. The color of each cell shows the \colorbox[HTML]{F09BA0}{best} and the \colorbox[HTML]{FCCF93}{second best}.}
\resizebox{1.0\linewidth}{!}{
\begin{tabular}{l|ccc|ccc|ccc|ccc|ccc}
\toprule
Scene       & \multicolumn{3}{c}{broom2}                                                                    & \multicolumn{3}{c}{vrig-3dprinter}                                                            & \multicolumn{3}{c}{vrig-chicken}                                                              & \multicolumn{3}{c}{vrig-peel-banana}                                                          & \multicolumn{3}{c}{Aveage}                                                                    \\
Method      & SSIM↑                         & PSNR↑                         & LPIPS↓                        & SSIM↑                         & PSNR↑                         & LPIPS↓                        & SSIM↑                         & PSNR↑                         & LPIPS↓                        & SSIM↑                         & PSNR↑                         & LPIPS↓                        & SSIM↑                         & PSNR↑                         & LPIPS↓                        \\
\midrule
HyperNeRF~\cite{park2021hypernerf}    & 0.210                         & 19.51                         & —                             & 0.635                         & 20.04                         & —                             & 0.828                         & 27.46                         & —                             & 0.719                         & 22.15                         & —                             & 0.598                         & 22.29                         & —                             \\
D3DGS~\cite{yang2024deformable}       & 0.269                         & 19.99                         & 0.700                         & 0.656                         & 20.71                         & 0.277                         & 0.640                         & 22.77                         & 0.363                         & 0.853                         & 25.95                         & \cellcolor[HTML]{F09BA0}0.155 & 0.605                         & 22.35                         & 0.374                         \\
MotionGS~\cite{zhu2024motiongs}    & 0.380                         & \cellcolor[HTML]{FCCF93}22.30 & —                             & 0.710                         & 21.80                         & —                             & 0.790                         & 26.80                         & —                             & 0.690                         & 28.20                         & —                             & 0.643                         & 24.78                         & —                             \\
MoDec-GS~\cite{kwak2025modecgs}    & 0.303                         & 21.04                         & 0.666                         & 0.706                         & 22.00                         & 0.265                         & 0.834                         & 28.77                         & 0.197                         & 0.873                         & 28.25                         & 0.173                         & 0.679                         & 25.02                         & 0.325                         \\
4DGaussians~\cite{wu20244d}  & 0.366                         & 22.01                         & 0.557                         & 0.705                         & 21.98                         & 0.327                         & 0.806                         & 28.49                         & 0.297                         & 0.847                         & 27.73                         & 0.204                         & 0.681                         & 25.05                         & 0.346                         \\
ED3DGS~\cite{bae2024per}       & 0.371                         & 21.84                         & 0.531                         & 0.715                         & 22.34                         & 0.294                         & 0.836                         & 28.75                         & \cellcolor[HTML]{FCCF93}0.185 & 0.867                         & \cellcolor[HTML]{FCCF93}28.80 & 0.178                         & 0.697                         & 25.43                         & 0.297                         \\
Grid4D~\cite{xu2024grid4d}       & \cellcolor[HTML]{FCCF93}0.414 & 21.78                         & \cellcolor[HTML]{FCCF93}0.423 & \cellcolor[HTML]{FCCF93}0.723 & \cellcolor[HTML]{FCCF93}22.36 & \cellcolor[HTML]{F09BA0}0.245 & \cellcolor[HTML]{FCCF93}0.848 & \cellcolor[HTML]{FCCF93}29.27 & 0.199                         & \cellcolor[HTML]{FCCF93}0.875 & 28.44                         & \cellcolor[HTML]{FCCF93}0.167 & \cellcolor[HTML]{FCCF93}0.715 & \cellcolor[HTML]{FCCF93}25.46 & \cellcolor[HTML]{FCCF93}0.259 \\
Ours        & \cellcolor[HTML]{F09BA0}0.422 & \cellcolor[HTML]{F09BA0}22.36 & \cellcolor[HTML]{F09BA0}0.413 & \cellcolor[HTML]{F09BA0}0.724 & \cellcolor[HTML]{F09BA0}22.56 & \cellcolor[HTML]{FCCF93}0.264 & \cellcolor[HTML]{F09BA0}0.858 & \cellcolor[HTML]{F09BA0}29.57 & \cellcolor[HTML]{F09BA0}0.166 & \cellcolor[HTML]{F09BA0}0.876 & \cellcolor[HTML]{F09BA0}28.81 & 0.169                         & \cellcolor[HTML]{F09BA0}0.720 & \cellcolor[HTML]{F09BA0}25.83 & \cellcolor[HTML]{F09BA0}0.253 \\
\bottomrule
\end{tabular}}
\label{vrig}
\end{table*}

\begin{table}[t]
\caption{\textbf{Quantitative comparison to previous methods on D-NeRF ~\cite{pumarola2021d} dataset.} The color of each cell shows the \colorbox[HTML]{F09BA0}{best} and the \colorbox[HTML]{FCCF93}{second best}. More detail results can be found in supplementary material.}
\resizebox{1.0\linewidth}{!}{
\begin{tabular}{l|ccc}
\toprule
Method      & SSIM↑                         & PSNR↑                         & LPIPS↓                        \\ \midrule
3DGS        ~\cite{kerbl20233d} & 0.930                         & 23.40                         & 0.077                         \\
K-Planes    ~\cite{fridovich2023k} & 0.970                         & 31.41                         & 0.047                         \\
HexPlane    ~\cite{cao2023hexplane} & 0.972                         & 31.92                         & 0.038                         \\
4DGaussians ~\cite{wu20244d} & 0.985                         & 35.32                         & 0.021                         \\
D3DGS ~\cite{yang2024deformable}  & 0.991                         & 40.08                         & 0.013                         \\
SC-GS       ~\cite{huang2024sc} & \cellcolor[HTML]{FCCF93}0.993 & 41.66                         & 0.009                         \\
Grid4D      ~\cite{xu2024grid4d} & \cellcolor[HTML]{F09BA0}0.994 & \cellcolor[HTML]{FCCF93}41.99 & \cellcolor[HTML]{FCCF93}0.008 \\
\midrule
Ours        & \cellcolor[HTML]{F09BA0}0.994 & \cellcolor[HTML]{F09BA0}42.17 & \cellcolor[HTML]{F09BA0}0.007 \\ \bottomrule
\end{tabular}}
\label{derf3}
\end{table}

\begin{figure*}[t]
\centerline{\includegraphics[width=0.92\textwidth]{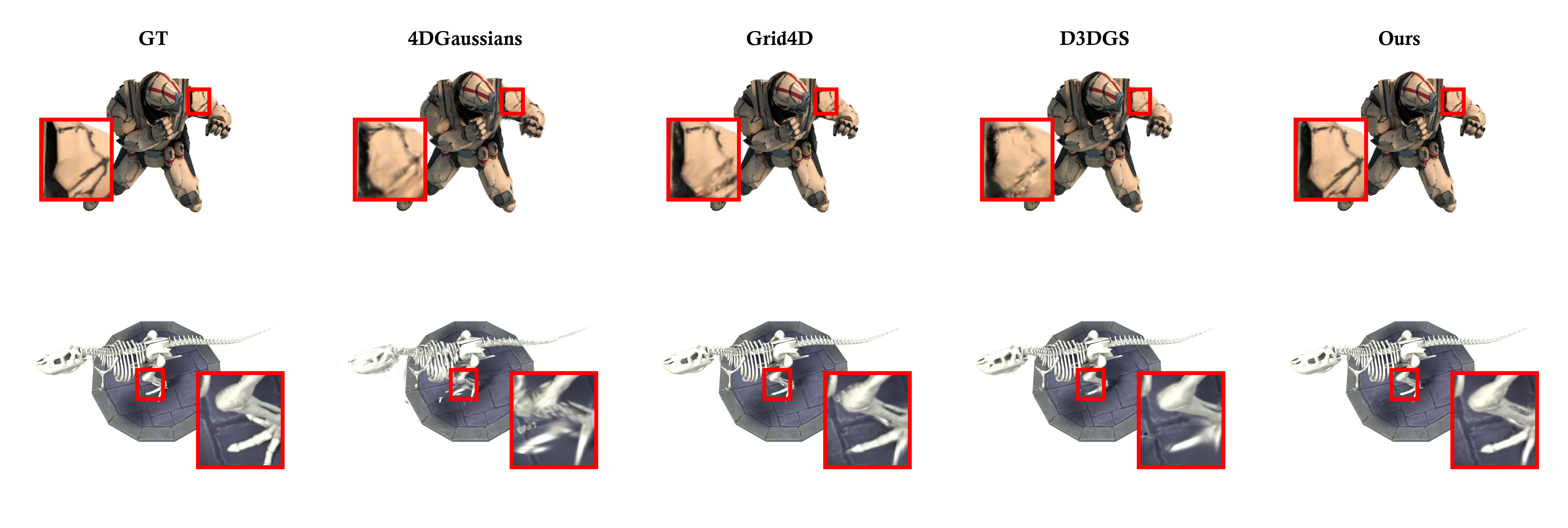}}
\caption{\textbf{Comparison Results.} Visual differences are highlighted with red insets for better clarity. Our approach consistently outperforms other models on D-NeRF ~\cite{pumarola2021d} dataset, demonstrating clear advantages in challenging scenarios such as thin geometries and fine-scale details. Best viewed in color.}
\label{fig2}
\end{figure*}

\section{Experiment}

\subsection{Experiment Setup}

We evaluate our proposed method using three widely recognized datasets, comprising two real-world datasets and one synthetic dataset. 
The Neu3D ~\cite{li2022neural} dataset is a real-world collection that features multiple static cameras and includes between 18 to 21 multi-view videos. 
We generate 300 frames for each video and initial point clouds for each scene following 4DGaussians ~\cite{wu20244d}. 
HyperNeRF ~\cite{park2021hypernerf} is a real-world dataset that captures continuous views with intricate topological variations at each timestamp within a dynamic scene.
In our experiment, we utilized the “vrig” subset, which was captured using stereo cameras, training the model with data from one camera while validating it with data from the other. 
The D-NeRF ~\cite{pumarola2021d} dataset serves as a synthetic dataset tailored for monocular scenes, with each scene comprising between 50 to 200 frames. 
Due to discrepancies between the training and testing scenarios presented in the Lego subset of the D-NeRF ~\cite{pumarola2021d} dataset, we excluded it from our experimental analysis.

\begin{table}[]
\caption{\textbf{Quantitative comparison to previous methods on Neu3D ~\cite{li2022neural} dataset.} Color of each cell shows the \colorbox[HTML]{F09BA0}{best} and the \colorbox[HTML]{FCCF93}{second best}. We show the average results of all scenes. More detail results can be found in supplementary material.}
\resizebox{1.0\linewidth}{!}{
\begin{tabular}{l|ccc}
\toprule
Method      & SSIM↑                         & PSNR↑                         & LPIPS↓                        \\ \midrule
4DGaussians ~\cite{wu20244d} & 0.935                         & 30.36                         & 0.152                         \\
Grid4D ~\cite{xu2024grid4d}      & 0.934                         & 30.50                         & 0.147                         \\
Spacetime ~\cite{li2024spacetime}   & \cellcolor[HTML]{F09BA0}0.944 & 31.46                         & 0.142                         \\
ED3DGS ~\cite{bae2024per}      & \cellcolor[HTML]{FCCF93}0.943 & \cellcolor[HTML]{FCCF93}31.92 & \cellcolor[HTML]{FCCF93}0.139 \\
\midrule
Ours        & \cellcolor[HTML]{F09BA0}0.944 & \cellcolor[HTML]{F09BA0}32.12 & \cellcolor[HTML]{F09BA0}0.134 \\ \bottomrule
\end{tabular}}
\label{d3dv2}
\end{table}

\begin{table}[t]
\centering
\caption{\textbf{Ablation Results on Neu3D ~\cite{li2022neural} dataset.} The Adaptive Gaussian Split Strategy helps reduce the number of Gaussians. }
\resizebox{0.8\linewidth}{!}{
\begin{tabular}{l|c}
\toprule
Method & Gaussian Counts \\
\midrule
w/o adaptive split strategy   & 589k
      \\
Ours  & 433k \\ \bottomrule    
\end{tabular}}
\label{knn}
\end{table}

\begin{figure*}
\centerline{\includegraphics[width=0.92\textwidth]{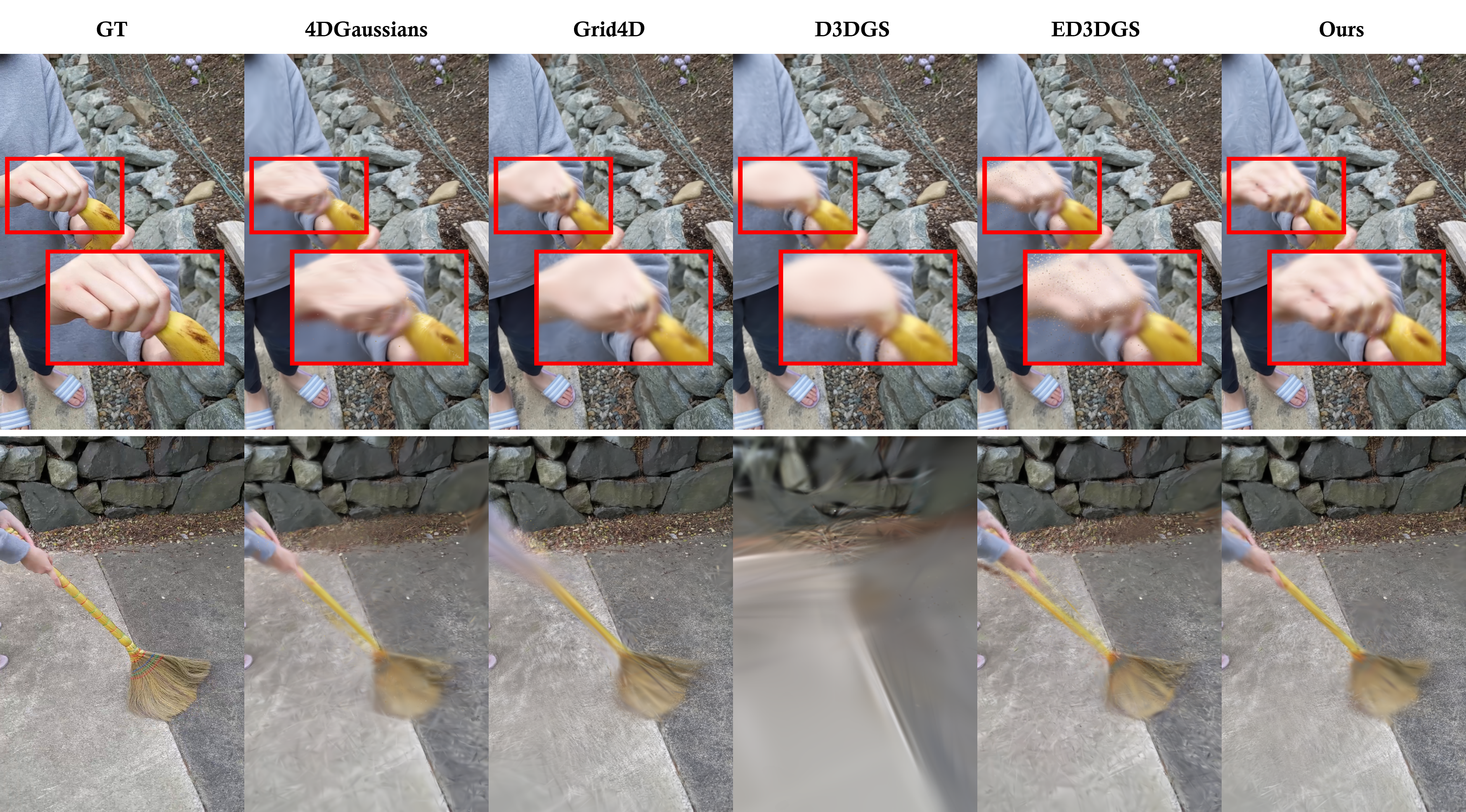}}
\caption{\textbf{Comparison Results.} Our approach consistently outperforms other models on HyperNeRF~\cite{park2021hypernerf} dataset, demonstrating clear advantages in challenging scenarios. Best viewed in color.}
\label{fig3}
\end{figure*}

\subsection{Comparisons}

On the Neu3D dataset, our approach demonstrates exceptional proficiency as shown in Tab. \ref{d3dv2}. 
The primary challenge here lies in accurately modeling intricate, often non-rigid, temporal dynamics while simultaneously reconstructing high-fidelity static scene geometry from these fixed perspectives. 
Our method excels in generating temporally coherent motion representations and preserving sharp geometric details throughout the sequences, effectively disentangling dynamic elements from the static background. 
In contrast, competing methods frequently struggle to maintain long-term temporal consistency across the multiple views, often exhibiting noticeable motion blur, 
particularly during complex actions or over extended durations.

The HyperNeRF “vrig” subset introduces a distinct set of demanding conditions. 
This dataset tests the model’s ability to handle complex motion while maintaining consistency across stereo viewpoints and adapt to evolving scene topology. 
Our method showcases remarkable resilience and adaptability in handling these extreme deformations and effectively leverages the stereo information, generalizing robustly across the viewpoints even when trained on one and validated on the other as shown in Tab. \ref{vrig} and Fig. \ref{fig3}. 
It consistently reconstructs intricate topological changes with greater accuracy and fewer visual artifacts or geometric distortions compared to existing approaches. 

Furthermore, evaluation on the synthetic D-NeRF dataset underscores our method’s inherent strength in inferring coherent 3D structure and plausible motion from limited input as shown in Tab. \ref{derf3} and Fig. \ref{fig2}. 
Reconstructing dynamic 3D geometry from a single, potentially moving, camera viewpoint over time presents profound depth ambiguities and necessitates strong priors and temporal reasoning. 
Despite this inherent ill-posedness and the scarcity of explicit geometric cues, our approach generates remarkably temporally stable and geometrically plausible reconstructions. 
Consequently, it significantly outperforms baseline methods, which, under these monocular constraints, often exhibit pronounced depth inaccuracies that betray instabilities in their representation.

Across this diverse range of evaluated datasets, our method consistently achieves a marked superiority in performance. 
This advantage is evident in both the final reconstruction fidelity and the accurate, coherent capture of dynamic motion, ranging from subtle deformations to large-scale topological changes. 
This consistent success across varied and demanding conditions robustly validates the effectiveness, versatility, and broad applicability of our proposed framework for dynamic scene reconstruction.

\begin{table}[]
\caption{\textbf{Ablation evaluation on Neu3D ~\cite{li2022neural} dataset.} The color of each cell shows the \colorbox[HTML]{F09BA0}{best}.}
\resizebox{1.0\linewidth}{!}{
\begin{tabular}{l|ccc}
\toprule
Method     & SSIM↑                         & PSNR↑                         & LPIPS↓                        \\ \midrule
w/o Laplacian module        & 0.938                         & 31.64                         & 0.149                         \\
w/o dynamic attribute       & 0.938                         & 31.72                         & 0.147                         \\
w/o adaptive split strategy & 0.943  & 31.96                          & \cellcolor[HTML]{F09BA0} 0.133                         \\
w/o $\mathcal{L}_{\text{lap}}$                    & 0.939 & 31.70 & 0.148 \\
\midrule
Ours                        & \cellcolor[HTML]{F09BA0}0.944                         & \cellcolor[HTML]{F09BA0}32.12                         & 0.134                         \\ \bottomrule
\end{tabular}}
\label{ablation}
\end{table}

\subsection{Ablation Study and Analysis}
\begin{figure}
\centerline{\includegraphics[width=0.45\textwidth]{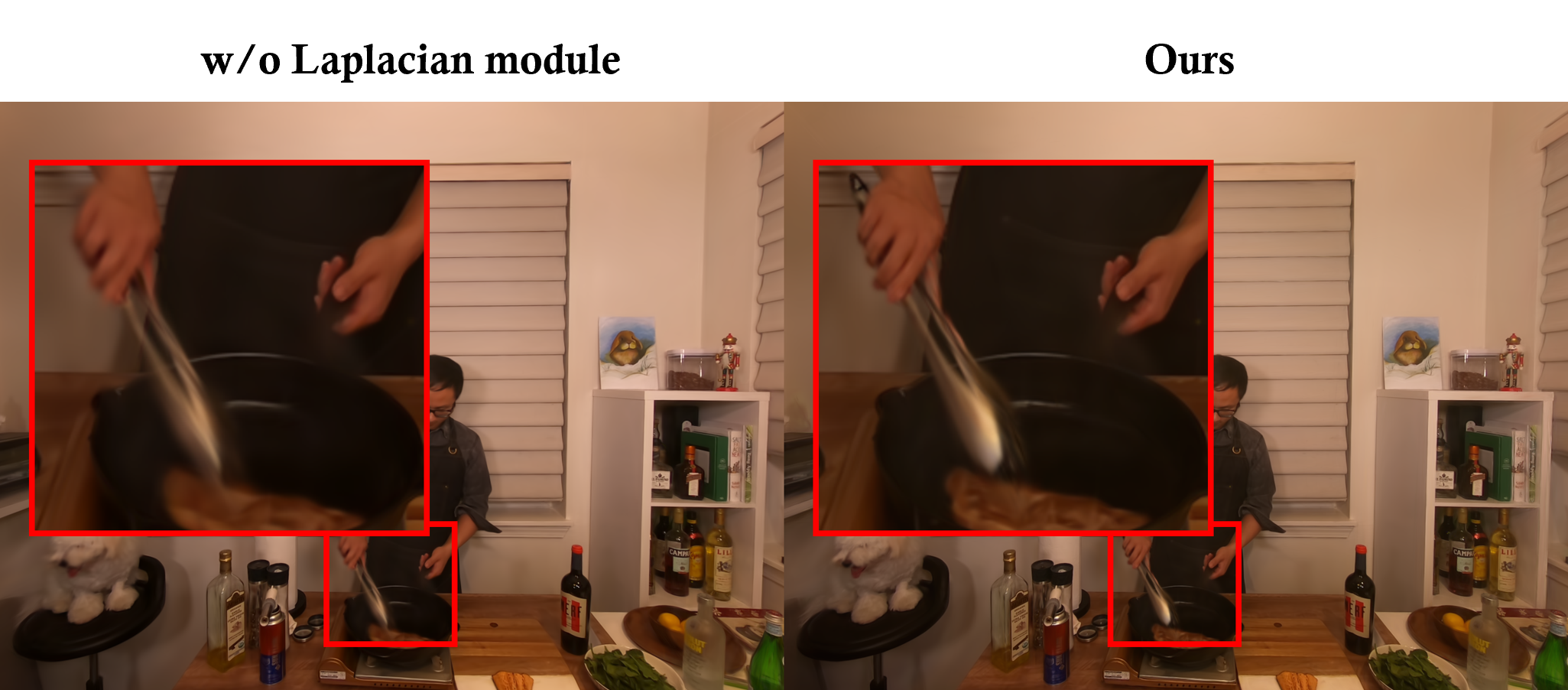}}
\caption{\textbf{Ablation Results.} Replacing Laplacian-based motion prediction leads to poorer performance.}
\label{fig4}
\end{figure}

To validate the effectiveness of each component within our framework, we conduct comprehensive ablation studies to validate the necessity of each component as shown in Tab. \ref{knn}, Tab. \ref{ablation} and Fig. \ref{fig4}.

\subsubsection{Effect of Laplacian-Based Motion Prediction}

Replacing this module with a defrom MLP leads to poorer performance, especially in scenes with diverse motion patterns or objects of varying sizes evolving over time. 
Dynamic scenes inherently possess variations across multiple spatial and temporal scales. 
This module is designed to capture these hierarchies effectively. 
It allows the model to represent fine details of motion trajectories while enabling the modeling of slow, gradual changes. 
By processing information hierarchically, it ensures consistent and accurate representation of scene dynamics across different frequency.

\subsubsection{Effect of Adaptive Gaussian Split Strategy}

Removing this component and reverting to original 3DGS split strategy results in a drop in reconstruction. 
Crucially, this component offers a dual benefit. 
Firstly, it enhances reconstruction quality by intelligently allocating Gaussian primitives. 
It adaptively densifies regions with high dynamics while pruning redundant or insignificant Gaussians. 
This leads to a more accurate representation of the scene and results in a more compact set of Gaussians compared to non-adaptive methods while achieving similar quality as shown in Tab.~\ref{knn}.

\subsubsection{Effect of Laplacian Pyramid Loss}
When this loss is removed on the rendered image, we observe a noticeable degradation in reconstruction quality.
The Laplacian pyramid loss decomposes the reconstruction error across multiple frequency bands by comparing the Laplacian pyramids of the rendered and ground truth images.
This loss function proves essential because it enforces structural consistency across different scales, effectively preserving fine details that would otherwise be lost in single-scale supervision.

\subsubsection{Effect of Gaussian dynamics attribute}

Compared with our full approach, removing the Gaussian dynamics attribute leads to poorer performance. 
This performance difference underscores the importance of embedding dedicated dynamic attributes within the Gaussians themselves. 
By incorporating these dynamic attribute, our method allows each Gaussian to better adapt its shape and orientation according to its specific local dynamics, effectively capturing details and mitigating the feature collision issues inherent in relying solely on lower-rank spatio-temporal grids.

\section{Conclusion}

In this paper, we present a novel approach for dynamic 3DGS that addresses the challenges of anisotropic spatio-temporal sampling through a hybrid explicit-implicit encoding framework. 
We introduced three key innovations: Firstly, a hybrid motion representation combining multi-scale Hash encoding with a Laplacian-based dynamic module, effectively decoupling different motion frequencies from complex deformation details.
Secondly, an enhanced Gaussian dynamics attribute that compensates for highly dynamic areas induced by geometric deformation. 
Thirdly, an adaptive Gaussian split strategy guided by KDTree-based analysis, which automatically adjusts dynamic primitive density and anisotropy.
This work advance the state-of-the-art in dynamic scene modeling by bridging the gap between explicit representations and spectral analysis, with potential applications in VR/AR and scene reconstruction.

\bibliography{aaai2026}

\end{document}